\documentclass[twocolumn,conference]{IEEEtran}
\usepackage[T1]{fontenc}
\usepackage[latin9]{inputenc}
\PassOptionsToPackage{ruled, linesnumbered}{algorithm2e}
\usepackage{algorithm2e}
\usepackage{amsmath}
\usepackage{amsthm}
\usepackage{amssymb}
\usepackage{graphicx}
\usepackage[unicode=true,
 bookmarks=true,bookmarksnumbered=true,bookmarksopen=true,bookmarksopenlevel=1,
 breaklinks=false,pdfborder={0 0 0},pdfborderstyle={},backref=false,colorlinks=false]
 {hyperref}
\hypersetup{pdftitle={Your Title},
 pdfauthor={Your Name},
 pdfpagelayout=OneColumn, pdfnewwindow=true, pdfstartview=XYZ, plainpages=false}

\makeatletter
\theoremstyle{plain}
\ifx\thechapter\undefined
	\newtheorem{thm}{\protect\theoremname}
\else
	\newtheorem{thm}{\protect\theoremname}[chapter]
\fi

\usepackage{algorithmic}
\usepackage[caption=false,font=footnotesize]{subfig}
\usepackage{amsmath,balance}
\usepackage{cite}
\usepackage{pgfplots}
\usepackage{pgfplotstable}
\usepackage{tikz}
\usepackage{graphicx}

\tikzset{mark options={line width=1pt,solid}}%
\definecolor{myred}{rgb}{0.95,0.00000,0}%
\definecolor{mygray}{rgb}{0.9,0.9,0.9}%
\definecolor{mygreen}{rgb}{0,0.5,0}%
\definecolor{myblue}{rgb}{0.4,0.42,0.8}%
\definecolor{myblack}{rgb}{0.2,0.2,0.2}%
\definecolor{mypurple}{rgb}{0.45,0,0.9}%
\definecolor{mywhite}{rgb}{1,1,1}%
\definecolor{mymagenta}{rgb}{0.95,0,0.95}%

\usepackage{color}
\DeclareMathOperator{\trace}{Tr}
\DeclareMathOperator{\maximize}{maximize}

\DeclareMathOperator{\st}{subject \, to}
\DeclareMathOperator{\diag}{diag}
\DeclareMathOperator{\vecd}{vec_d}

\newcommand{\herm}{^{{\dagger}}}
\newcommand{\trans}{^{\mbox{\scriptsize T}}}

\allowdisplaybreaks
\usepackage[left=0.6in, right=0.6in, top=0.7in, bottom=1in,centering]{geometry}

\makeatother

\providecommand{\theoremname}{Theorem}

\begin{document}
\title{On the Achievable Rate of IRS-Assisted Multigroup Multicast Systems}
\author{\IEEEauthorblockN{Muhammad~Farooq\IEEEauthorrefmark{1}, Vaibhav~Kumar\IEEEauthorrefmark{1},
Markku Juntti\IEEEauthorrefmark{2}, and Le-Nam~Tran\IEEEauthorrefmark{1}}\IEEEauthorblockA{\IEEEauthorrefmark{1}School of Electrical and Electronic Engineering,
University College Dublin, Belfield, Dublin 4, Ireland\\
\IEEEauthorrefmark{2}Centre for Wireless Communications, University
of Oulu, P.O. Box 4500, FI-90014, Finland\\
Email: muhammad.farooq@ucdconnect.ie; vaibhav.kumar@ieee.org; markku.juntti@oulu.fi;
nam.tran@ucd.ie}}

\maketitle
{\let\thefootnote\relax\footnotetext{This publication has emanated from research conducted with the financial support of Science Foundation Ireland (SFI) and is co-funded under the European Regional Development Fund under Grant Number 17/CDA/4786.}}
\begin{abstract}
Intelligent reflecting surfaces (IRSs) have shown huge advantages
in many potential use cases and thus have been considered a promising
candidate for next-generation wireless systems. In this paper, we
consider an IRS-assisted multigroup multicast (IRS-MGMC) system in
a multiple-input single-output (MISO) scenario, for which the related
existing literature is rather limited. In particular, we aim to jointly
design the transmit beamformers and IRS phase shifts to maximize the
sum rate of the system under consideration. In order to obtain a numerically
efficient solution to the formulated non-convex optimization problem,
we propose an \textit{alternating projected gradient} (APG) method
where each iteration admits a closed-form and is shown to be superior
to a known solution that is derived from the majorization-minimization
(MM) method in terms of both achievable sum rate and required complexity,
i.e., run time. In particular, we show that the complexity of the
proposed APG method grows \textit{linearly} with the number of IRS
tiles, while that of the known solution in comparison grows with the
third power of the number of IRS tiles. The numerical results reported
in this paper extend our understanding on the achievable rates of
large-scale IRS-assisted multigroup multicast systems.
\end{abstract}

\begin{IEEEkeywords}
Intelligent reflecting surface, multigroup multicast, alternating
projected gradient, multiple-input single-output.
\end{IEEEkeywords}

\section{Introduction}

With the advent of new wireless communication services and an unprecedented
increase in the number of wireless devices, the problem of supporting
massive connectivity in the very-congested sub-6 GHz spectrum band
has become a challenging problem for beyond fifth generation (B5G)/sixth
generation (6G) wireless standards. In this context, services like
audio/video streaming are supported via the evolved multimedia broadcast/multicast
service (eMBMS) in the third-generation partnership project (3GPP)~\cite{eMBMS,5GNR}.
In such a media streaming scenario, multiple users in a group are
served with the same content, whereas the content delivered among
the groups is mutually independent. This indeed has motivated several
pioneer works on physical layer multicasting~\cite{PhyLayerMulticast-Luo,MulticastQoS-Luo}.

On the other hand, intelligent reflecting surfaces (IRSs) are being
envisioned as a disruptive technology to enhance the performance of
next-generation wireless~\cite{IRSmag21}. These IRSs are capable
of steering the incident radio waves in the desired direction to improve
a certain performance measure. Therefore, optimal system design for
IRS-assisted physical layer multicasting has naturally gained increasing
interest in recent years. Some of the recent works on IRS-assisted
\textit{single-group} multicast system design include~\cite{IRS-SGMC-TangTVT21,IRS-SGMC-TangTWC21,IRS-SGMC-Zhang21},
whereas those on IRS-assisted \textit{multigroup} multicast (IRS-MGMC)
system design include~\cite{Zhou20IRS,IRS-MGMC-VTC20,IRS-MGMC-SymRadioVTC21}.

In particular, Zou \emph{et al.} in~\cite{Zhou20IRS} considered
the problem of sum rate maximization in a multiple-input single-output
(MISO) IRS-MGMC system, where they aim to jointly design the optimal
transmit beamforming vectors and IRS phase shifts via alternating
optimization (AO). More specifically, they proposed two different
optimization schemes based on majorization-minimization (MM). The
first one involved second-order cone programming (SOCP) which requires
high complexity, while the second scheme was based on a smoothing
technique and admitted closed-form expressions, and thus had lower
complexity. We refer to the latter as the MM method for short in the
rest of the paper. The problem of transmit power minimization in an
MISO IRS-MGMC system subject to some quality-of-service (QoS) constraints
was presented in~\cite{IRS-MGMC-VTC20}, where the optimal transmit
beamformers and IRS phase shifts were obtained using a difference-of-convex
(DC) algorithm in conjunction with the AO. Similarly, in~\cite{IRS-MGMC-SymRadioVTC21},
the problem of transmit power minimization subject to multiple QoS
constraints in an IRS-MGMC symbiotic radio was considered, where optimal
transmit and reflect structures were obtained using AO, quadratic
transform, and semidefinite programming (SDP).

It is worth mentioning that in a practical system, the number of passive
reflecting tiles in the IRS will be in the order of a few hundred,
if not thousands~\cite{IRS_Prototype}. This massive number of reflecting
elements will likely contribute the most toward the total complexity
of any optimal system design. Therefore, for the IRS-MGMC sum rate
maximization problem, the MM method in~\cite[Algorithm~2]{Zhou20IRS},
whose complexity has a cubic growth with the number of IRS tiles,
is still impractical for large systems. Also, as shall be seen later,
the MM method in~\cite{Zhou20IRS} does not perform well in the high
power regime since the derived bounds were not tight in this case.
Due to the shortcomings of the previous studies, in this paper, we
propose a first-order method based on \textit{alternating projected
gradient} (APG), which results in superior performance and lower complexity
compared to that of the MM method in~\cite{Zhou20IRS}. Our main
contributions in this paper are as follows:
\begin{enumerate}
\item[(i)]  We propose a low-complexity APG algorithm to obtain optimal transmit
and reflect structures for sum rate maximization in an MISO IRS-MGMC
system, which is shown to outperform the benchmark MM method, i.e.
~\cite[Algorithm~2]{Zhou20IRS}.
\item[(ii)]  We provide a detailed complexity analysis of the proposed APG algorithm
which confirms that the per-iteration complexity of our proposed algorithm
\emph{grows linearly} with the number of IRS tiles, which is significantly
less than that of the benchmark MM method. In particular, our implementation
indicates that the proposed APG method is $1000\times$ faster when
the number of reflecting elements is $400$.
\item[(iii)]  Extensive numerical results are provided to evaluate the performance
of our proposed algorithm, and to analyze the effect of different
system parameters of interest on the system performance, showing that
the IRS can improve significantly the achievable sum rate in multigroup
multicasting scenarios. Specifically, the gain of the proposed solution
is up to $50\%$ compared to the MM method.
\end{enumerate}

\paragraph*{Notations}

Bold uppercase and lowercase letters are used to denote matrices and
vectors, respectively. The Hermitian, transpose and Euclidean norm
of a vector $\mathbf{x}$ are, respectively, denoted by $\mathbf{x}\herm$,
$\mathbf{x}\trans$ and $\Vert\mathbf{x}\Vert$. The absolute value
of a complex number $x$ is denoted by $|x|$. The vector space of
all $M\times N$ complex-valued matrices is denoted by $\mathbb{C}^{M\times N}$.
By $\diag(\mathbf{x})$, we represent a square diagonal matrix whose
main diagonal consists of the elements of vector $\mathbf{x}$, whereas,
$\vecd(\mathbf{X})$ represents the column vector whose elements are
the same as that of the main diagonal of the matrix $\mathbf{X}$.
The operation $\ln(\cdot)$ represents the natural logarithm of the
argument. The trace of the matrix $\mathbf{X}$ is denoted by $\trace(\mathbf{X})$.
Euclidean projection of a given vector $\mathbf{x}$ onto the set
$\mathcal{X}$ is defined as $\Pi_{\mathcal{X}}(\mathbf{x})\triangleq\min_{\hat{\mathbf{x}}\in\mathcal{X}}\Vert\hat{\mathbf{x}}-\mathbf{x}\Vert$.
The expected value of a random variable is denoted by $\mathbb{E}\{\cdot\}$.
By $\mathcal{O}(\cdot)$, we denote the Landau symbol. Finally, we
denote by $\nabla_{\mathbf{x}}f(\cdot)$ the complex gradient of $f(\cdot)$
with respect to $\mathbf{x}^{\ast}$, i.e., $\nabla_{\mathbf{x}}f(\cdot)\equiv\frac{\partial}{\partial\mathbf{x}^{\ast}}f(\cdot)=\frac{1}{2}\Bigl(\frac{\partial f(\cdot)}{\partial\Re(\mathbf{x})}+j\frac{\partial f(\cdot)}{\partial\Im(\mathbf{x})}\Bigr)$
\cite[Chap. 3]{DerivativeBook}.

\section{\label{sec:System-Model}System Model and Problem Formulation}

\begin{figure}[t]
\begin{centering}
\includegraphics[width=0.8\columnwidth]{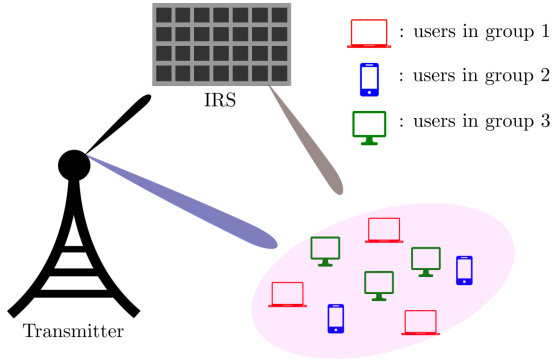}
\par\end{centering}
\caption{An example IRS-assisted multigroup multicast system model with $K=8$,
$G=3$, $K_{1}=K_{3}=3$, and $K_{2}=2$.}
\label{fig:SysMod}
\end{figure}

Consider an IRS-assisted multigroup multicast system shown in~Fig.~\ref{fig:SysMod},
consisting of one transmitter, one (passive) IRS, and $K$ users.
We assume that the transmitter is equipped with $N$ antennas, the
IRS consists of $M$ low-cost passive reflecting tiles, and all of
the users are single-antenna nodes. The users are divided into $G$
groups, where the set of groups is denoted by $\mathcal{G}\triangleq\{1,2,\ldots,G\}$.
We denote by $\mathrm{U}_{k,g}$ the $k$-th user in the $g$-th group.
The set of users in each group $g\in\mathcal{G}$ is denoted by $\mathcal{U}_{g}\triangleq\{\mathrm{U}_{k,g}\mid k\in\mathcal{K}_{g}\}$
where $\mathcal{K}_{g}\triangleq\{1,2,\ldots,K_{g}\}$ and $K_{g}$
denotes the number of users in the $g$-th group. We assume that all
of the users in one group receive the same message from the transmitter,
while the message intended for each group is independent from those
for the other groups. In this paper we adopt a beamforming approach
for each group. Thus, the signal at the transmitter is therefore given
by
\begin{equation}
\mathbf{x}=\sum\nolimits _{g\in\mathcal{G}}\mathbf{f}_{g}v_{g},\label{TxSignal}
\end{equation}
where $\mathbf{f}_{g}\in\mathbb{C}^{N\times1}$ is the beamforming
vector for the $g$-th group and $v_{g}$ is the information-bearing
Gaussian symbol intended for the $g$-th group such that $\mathbb{E}\{|v_{g}|^{2}\}=1$.
We define $\mathbf{f}\triangleq[\mathbf{f}_{1}\trans,\mathbf{f}_{2}\trans,\ldots,\mathbf{f}_{G}\trans]\trans\in\mathbb{C}^{NG\times1}$
to be the vector stacking the beamforming vectors of all of the groups.
We assume that the instantaneous channel state information (CSI) is
perfectly known at all of the nodes\footnote{A similar assumption on the perfect CSI availability was considered
in~\cite{IRS-SGMC-TangTVT21,Zhou20IRS,IRS-MGMC-VTC20,IRS-MGMC-SymRadioVTC21,StefanMIMOBC}.
Also, since in this paper our main motivation is to provide an efficient
low-complexity algorithm for the sum rate maximization problem in
an IRS-MGMC system, the problem of channel estimation is out of scope
of this paper.}. Let the transmitter-IRS, transmitter-$\mathrm{U}_{k,g}$, and IRS-$\mathrm{U}_{k,g}$
links are denoted by $\mathbf{H}_{\mathrm{ts}}\in\mathbb{C}^{M\times N}$,
$\mathbf{h}_{k,g}\in\mathbb{C}^{1\times N}$, and $\hat{\mathbf{h}}_{k,g}\in\mathbb{C}^{1\times M}$,
respectively. Then, the received signal at $\mathrm{U}_{k,g}$ is
expressed as
\begin{equation}
y_{k,g}=\big(\mathbf{h}_{k,g}+\hat{\mathbf{h}}_{k,g}\boldsymbol{\Theta}\mathbf{H}_{\mathrm{ts}}\big)\mathbf{x}+w_{k,g},\label{eq:RxSignal}
\end{equation}
where $w_{k,g}\sim\mathcal{CN}(0,\sigma^{2})$ is the additive white
Gaussian noise at $\mathrm{U}_{k,g}$, $\boldsymbol{\Theta}\triangleq\diag(\boldsymbol{\theta})$
with $\boldsymbol{\theta}\triangleq[\theta_{1},\theta_{2},\ldots,\theta_{M}]\trans\in\mathbb{C}^{M\times1}$
denoting the IRS phase shift vector, $\theta_{m}\triangleq\exp(j\phi_{m})$,
$\forall m\in\mathcal{M}\triangleq\{1,2,\ldots,M\}$, and $\phi_{m}\in[0,2\pi)$.
With a slight abuse of notation, in the sequel of the paper, we normalize
the involving channels appropriately with the noise power, i..e, $\mathbf{h}_{k,g}\leftarrow\mathbf{h}_{k,g}/\sigma$
and $\hat{\mathbf{h}}_{k,g}\leftarrow\hat{\mathbf{h}}_{k,g}/\sigma$,
and thus the resulting equivalent noise has a unit variance. In addition
to lightening the notations, this normalization step also has a numerical
benefit in the sense that we do not need to deal directly with extremely
small quantities such as the channels and the noise power themselves.
In this way, the achievable rate (in nats/s/Hz) at $\mathrm{U}_{k,g}$
is given by\footnote{For mathematical convenience, we express the achievable rate in nats/s/Hz.
However, the results in Section~\ref{sec:Numerical-Results} are
shown for achievable rate in bps/Hz, which is more commonly used.}
\begin{align}
R_{g,k}(\mathbf{f},\boldsymbol{\theta}) & =\ln\left(1+\frac{|\mathbf{z}_{g,k}\mathbf{f}_{g}|^{2}}{1+\sum_{\ell\in\mathcal{G}\setminus\{g\}}|\mathbf{z}_{k,g}\mathbf{f}_{\ell}|^{2}}\right),\label{eq:userrate}
\end{align}
where $\mathbf{z}_{k,g}\triangleq\mathbf{h}_{k,g}+\hat{\mathbf{h}}_{k,g}\boldsymbol{\Theta}\mathbf{H}_{\mathrm{ts}}$.
We emphasize that $\mathbf{z}_{k,g}$ depends on the phase shifts,
although the notation does not explicitly show this. Since all of
the users in the $g$-th group needs to correctly decode the intended
message, i.e., $v_{g}$, the achievable rate for the $g$-th group
is defined as
\begin{equation}
\mathcal{R}_{g}(\mathbf{f},\boldsymbol{\theta})=\min\nolimits _{k\in\mathcal{K}_{g}}\{R_{k,g}(\mathbf{f},\boldsymbol{\theta})\}.\label{eq:grouprate}
\end{equation}

In this paper, we aim to jointly design the optimal transmit beamformers
$(\mathbf{f}_{g},\forall g\in\mathcal{G})$ and IRS phase shifts $(\theta_{m},\forall m\in\mathcal{M})$
to maximize achievable sum rate of all of the groups. The optimization
problem can therefore be mathematically stated as 
\begin{subequations}
\label{eq:mainproblem}
\begin{align}
\underset{\mathbf{f},\boldsymbol{\theta}}{\maximize} & \ \big\{\mathcal{R}_{\mathrm{sum}}(\mathbf{f},\boldsymbol{\theta})\triangleq\sum\nolimits _{g\in\mathcal{G}}\mathcal{R}_{g}(\mathbf{f},\boldsymbol{\theta})\big\}\label{eq:systemrate}\\
\st & \ \Vert\mathbf{f}\Vert\leq\sqrt{P_{\mathrm{t}}},\label{eq:TPC}\\
 & \ |\theta_{m}|=1,\forall m\in\mathcal{M},\label{eq:UMCs}
\end{align}
\end{subequations}
where $P_{\mathrm{t}}$ is the maximum transmit power available at
the transmitter. The constraint in~\eqref{eq:TPC} denotes the transmit
power constraint at the transmitter, and those in~\eqref{eq:UMCs}
represent the unit-modulus constraints at the reflecting tiles\footnote{Many different IRS phase shift models are suggested in literature~\cite{OptMag,IRS_Prototype},
however, the unit-modulus model is the most frequently used one~\cite{IRSmag21,IRS-SGMC-TangTVT21,IRS-SGMC-TangTWC21,IRS-SGMC-Zhang21,Zhou20IRS,IRS-MGMC-VTC20,IRS-MGMC-SymRadioVTC21,StefanMIMOBC}.}. Note that the problem in~\eqref{eq:mainproblem} is non-convex
due to the coupling between the optimization variables $\mathbf{f}$
and $\boldsymbol{\theta}$ in~\eqref{eq:systemrate}, and the non-convex
constraints in~\eqref{eq:UMCs}. It is also noteworthy that~\eqref{eq:systemrate}
is non-differentiable due to the inherent piecewise minimization operator
in~\eqref{eq:grouprate}. These two challenges make it difficult
to develop efficient methods for solving \eqref{eq:mainproblem}.

Before describing our proposed solution in the next section, we provide
an important remark regarding an existing solution to~\eqref{eq:mainproblem},
proposed in~\cite[Algorithm~1]{Zhou20IRS} based on the MM method.
In particular, the authors first obtained a \textit{concave lower
bound} on~\eqref{eq:systemrate} when $\mathbf{f}$ or $\boldsymbol{\theta}$
are fixed and then maximize the derived lower bound. This process
is repeated alternatively between $\mathbf{f}$ and $\boldsymbol{\theta}$
until convergence. In particular, it was shown that the resulting
per-iteration complexity of this method is $\mathcal{O}(M^{3.5})$
\footnote{In a practical IRS-MGMC system, $M\gg\max\{N,K,G\}$, and therefore
we neglect the lower-order terms.}. Furthermore, in order to reduce the complexity, the authors in~\cite[Algorithm~2]{Zhou20IRS}
also proposed the MM method with the per-iteration complexity of $\mathcal{O}(K(M+1)^{3})$
based on a smoothing technique. From the complexity perspective, it
is obvious that both methods are still not practically appealing for
large $M$ which is expected to be the case for the IRS to have a
significant impact. Thus, developing a more efficient algorithm for
solving \eqref{eq:mainproblem} is still of huge interest.

\section{\label{sec:Proposed-Solution}Proposed Solution}

In this section, we propose a low-complexity algorithm to find a stationary
solution to the problem in~\eqref{eq:mainproblem}, and show that
the complexity of our proposed algorithm grows \textit{linearly} with
the number of tiles at the IRS.

\subsection{The Alternating Projected Gradient (APG) Algorithm}

First, to tackle the nonsmoothness of the objective, we apply a smoothing
technique introduced in \cite{Nesterov2005a}. We remark that although
a smoothing techniques is also used in~\cite{Zhou20IRS}, the difference
is that in this paper, we apply the smoothing on the true objective,
i.e., on~\eqref{eq:systemrate}, whereas in~\cite{Zhou20IRS}, the
same smoothing was applied to a concave lower-bound on~\eqref{eq:systemrate}.
As shall be demonstrated in the next section, it turns out that this
difference creates a huge impact on the achievable rate. Specifically,
a differentiable approximation of~\eqref{eq:systemrate} is found
as \cite{Nesterov2005a}
\begin{align}
\!\mathcal{R}_{\mathrm{sum}}(\mathbf{f},\boldsymbol{\theta})\! & \approx\!-\frac{1}{\tau}\sum\nolimits _{g\in\mathcal{G}}\ln\Big[\sum\nolimits _{k\in\mathcal{K}_{g}}\exp\big\{\!-\tau R_{k,g}(\mathbf{f},\boldsymbol{\theta})\big\}\Big]\nonumber \\
 & \triangleq\tilde{\mathcal{R}}_{\mathrm{sum}}(\mathbf{f},\boldsymbol{\theta}),\label{eq:approxsystemrate}
\end{align}
where $\tau>0$ is the smoothing parameter, which also determines
the accuracy of the approximation. Therefore, the problem in \eqref{eq:mainproblem}
is modified to the following problem
\begin{subequations}
\label{eq:approxproblem}
\begin{align}
\underset{\mathbf{f},\boldsymbol{\theta}}{\maximize} & \ \tilde{\mathcal{R}}_{\mathrm{sum}}(\mathbf{f},\boldsymbol{\theta}),\label{eq:approxObj}\\
\st & \ \eqref{eq:TPC},\eqref{eq:UMCs}.\nonumber 
\end{align}
\end{subequations}
For ease of exposition, we define the feasible set for $\mathbf{f}$
and $\boldsymbol{\theta}$ by $\mathcal{F}\triangleq\{\mathbf{f}\in\mathbb{C}^{NG\times1}\rvert\ \Vert\mathbf{f}\Vert\le\sqrt{P_{\mathrm{t}}}\}$
and $\varTheta=\{\boldsymbol{\theta}\in\mathbb{C}^{M\times1}|\ |\theta_{m}|=1,m\in\mathcal{M}\}$,
respectively. We remark that the sets $\mathcal{F}$ and $\varTheta$
are simple in the sense that their projection can be computed exactly
by closed-form. This fact indeed motivates us to adopt the alternating
projected gradient to solve \eqref{eq:mainproblem}. In this context,
we first provide the (complex-valued) gradient of $\tilde{\mathcal{R}}_{\mathrm{sum}}(\mathbf{f},\boldsymbol{\theta})$
with respect to (w.r.t.) $\mathbf{f}$. By definition, we can immediately
write $\nabla_{\mathbf{f}}\tilde{\mathcal{R}}_{\mathrm{sum}}(\mathbf{f},\boldsymbol{\theta})$
as 
\begin{align}
 & \nabla_{\mathbf{f}}\tilde{\mathcal{R}}_{\mathrm{sum}}(\mathbf{f},\boldsymbol{\theta})=\big[\big(\nabla_{\mathbf{f}_{1}}\tilde{\mathcal{R}}_{\mathrm{sum}}(\mathbf{f},\boldsymbol{\theta})\big)\trans,\nonumber \\
 & \qquad\big(\nabla_{\mathbf{f}_{2}}\tilde{\mathcal{R}}_{\mathrm{sum}}(\mathbf{f},\boldsymbol{\theta})\big)\trans,\ldots,\big(\nabla_{\mathbf{f}_{G}}\tilde{\mathcal{R}}_{\mathrm{sum}}(\mathbf{f},\boldsymbol{\theta})\big)\trans\big]\trans,\label{eq:gradFcomp}
\end{align}
where $\nabla_{\mathbf{f}_{i}}\tilde{\mathcal{R}}_{\mathrm{sum}}(\mathbf{f},\boldsymbol{\theta}),i\in\mathcal{G}$,
is provided by the following theorem.

\begin{algorithm}[t]
\caption{The proposed alternating projected gradient (APG) algorithm for solving~\eqref{eq:approxproblem}}

\label{alg:projGrad}

\KwIn{$\mathbf{f}^{(0)}$, $\boldsymbol{\theta}^{(0)}$, $\alpha_{\mathbf{f}}$,
$\alpha_{\boldsymbol{\theta}}$, $\tau>0$ }

\KwOut{$\mathbf{f}^{\star}$, $\boldsymbol{\theta}^{\star}$}

$n\leftarrow1$

\Repeat{convergence }{

$\!\!\mathbf{f}^{(n)}=\Pi_{\mathcal{F}}\big(\mathbf{f}^{(n-1)}+\alpha_{\mathbf{f}}\nabla_{\mathbf{f}}\tilde{\mathcal{R}}_{\mathrm{sum}}\big(\mathbf{f}^{(n-1)},\boldsymbol{\theta}^{(n-1)}\big)\big)$\;

$\boldsymbol{\theta}^{(n)}=\Pi_{\varTheta}\big(\boldsymbol{\theta}^{(n-1)}+\alpha_{\boldsymbol{\theta}}\nabla_{\boldsymbol{\theta}}\tilde{\mathcal{R}}_{\mathrm{sum}}\big(\mathbf{f}^{(n)},\boldsymbol{\theta}^{(n-1)}\big)\big)$\;

$\!\!$$n\leftarrow n+1$\;

}

$\mathbf{f}^{\star}\leftarrow\mathbf{f}^{(n)}$, $\boldsymbol{\theta}^{\star}\leftarrow\boldsymbol{\theta}^{(n)}$
\end{algorithm}

\begin{thm}
\label{thm:Grad-f}A closed-form expression for $\nabla_{\mathbf{f}_{i}}\tilde{\mathcal{R}}_{\mathrm{sum}}(\mathbf{f},\boldsymbol{\theta})$
is given by~\eqref{eq:gradFclosed}, shown at the top of the next
page, where $\nabla_{\mathbf{f}_{i}}R_{k,i}(\mathbf{f},\boldsymbol{\theta})=\frac{\mathbf{z}_{k,i}\herm\mathbf{z}_{k,i}\mathbf{f}_{i}}{1+\sum_{g\in\mathcal{G}}|\mathbf{z}_{k,i}\mathbf{f}_{g}|^{2}}$
and $\nabla_{\mathbf{f}_{i}}R_{k,\ell}(\mathbf{f},\boldsymbol{\theta})=\big[\big(1+\sum_{g\in\mathcal{G}}|\mathbf{z}_{k,\ell}\mathbf{f}_{g}|^{2}\big)^{-1}-\big(1+\sum_{\jmath\in\mathcal{G}\setminus\{\ell\}}|\mathbf{z}_{k,\ell}\mathbf{f}_{\jmath}|^{2}\big)^{-1}\big]\mathbf{z}_{k,\ell}\herm\mathbf{z}_{k,\ell}\mathbf{f}_{i}$.
\end{thm}
\begin{IEEEproof}
See Appendix~\ref{sec:Proof-1}.
\end{IEEEproof}
\begin{figure*}[t]
\small{
\begin{align}
\nabla_{\mathbf{f}_{i}}\tilde{\mathcal{R}}_{\mathrm{sum}}(\mathbf{f},\boldsymbol{\theta}) & =\frac{\sum_{k\in\mathcal{K}_{i}}\big[\exp\big\{-\tau R_{k,i}(\mathbf{f},\boldsymbol{\theta})\big\}\nabla_{\mathbf{f}_{i}}R_{k,i}(\mathbf{f},\boldsymbol{\theta})\big]}{\sum_{k\in\mathcal{K}_{i}}\exp\big\{-\tau R_{k,i}(\mathbf{f},\boldsymbol{\theta})\big\}}+\sum_{\ell\in\mathcal{G}\setminus\{i\}}\frac{\sum_{k\in\mathcal{K}_{\ell}}\big[\exp\big\{-\tau R_{k,\ell}(\mathbf{f},\boldsymbol{\theta})\big\}\nabla_{\mathbf{f}_{i}}R_{k,\ell}(\mathbf{f},\boldsymbol{\theta})\big]}{\sum_{k\in\mathcal{K}_{\ell}}\exp\big\{-\tau R_{k,\ell}(\mathbf{f},\boldsymbol{\theta})\big\}}.\label{eq:gradFclosed}
\end{align}
}
\end{figure*}
Next, we obtain a closed-form expression for the gradient of $\tilde{\mathcal{R}}_{\mathrm{sum}}(\mathbf{f},\boldsymbol{\theta})$
w.r.t. $\boldsymbol{\theta}$ as given below.
\begin{thm}
\label{thm:Grad-theta}A closed-form expression for $\nabla_{\boldsymbol{\theta}}\tilde{\mathcal{R}}_{\mathrm{sum}}(\mathbf{f},\boldsymbol{\theta})$
is given by~\eqref{eq:gradThetaClosed}, shown at the top of the
next page, where $\nabla_{\boldsymbol{\theta}}|\mathbf{z}_{k,g}\mathbf{f}_{\imath}|^{2}=\vecd\big\{\big(\hat{\mathbf{h}}_{k,g}\big)\herm\mathbf{z}_{k,g}\mathbf{f}_{\imath}\mathbf{f}_{\imath}\herm\mathbf{H}_{\mathrm{ts}}\herm\big\}$.
\end{thm}
\begin{IEEEproof}
See Appendix~\ref{sec:Proof-2}.
\end{IEEEproof}
\begin{figure*}[t]
\small{
\begin{align}
\nabla_{\boldsymbol{\theta}}\tilde{\mathcal{R}}_{\mathrm{sum}}(\mathbf{f},\boldsymbol{\theta}) & =\sum\nolimits _{g\in\mathcal{G}}\dfrac{\sum\nolimits _{k\in\mathcal{K}_{g}}\Bigg[\exp\big\{-\tau R_{k,g}(\mathbf{f},\boldsymbol{\theta})\big\}\Bigg\{\dfrac{\sum\nolimits _{\jmath\in\mathcal{G}}\nabla_{\boldsymbol{\theta}}|\mathbf{z}_{k,g}\mathbf{f}_{\jmath}|^{2}}{1+\sum\nolimits _{\jmath\in\mathcal{G}}|\mathbf{z}_{k,g}\mathbf{f}_{\jmath}|^{2}}-\dfrac{\sum\nolimits _{\ell\in\mathcal{G}\setminus\{\jmath\}}\nabla_{\boldsymbol{\theta}}|\mathbf{z}_{k,g}\mathbf{f}_{\ell}|^{2}}{1+\sum\nolimits _{\ell\in\mathcal{G}\setminus\{\jmath\}}|\mathbf{z}_{k,g}\mathbf{f}_{\ell}|^{2}}\Bigg\}\Bigg]}{\sum\nolimits _{k\in\mathcal{K}_{g}}\exp\big\{-\tau R_{k,g}(\mathbf{f},\boldsymbol{\theta})\big\}}.\label{eq:gradThetaClosed}
\end{align}
}

\hrulefill
\end{figure*}

\begin{figure*}[t]
\begin{minipage}[t]{0.32\linewidth}%
\begin{center}
\newcommand{\vasymptote}[2][]{
	\draw [densely dashed,#1] ({rel axis cs:0,0} -| {axis cs:#2,0}) -- ({rel axis cs:0,1} -| {axis cs:#2,0});
}
\pgfplotsset{every tick label/.append style={font=\large}}
\resizebox{0.98\columnwidth}{0.75\columnwidth}{
	\begin{tikzpicture}
	\begin{axis}[
	ymin=0,ymax=13,xmin=0,xmax=1500,
	xtick={0,500,1000,1500},
	ytick={0,2,4,6,8,10,12},
	grid=both,
	minor grid style={gray!25},
	major grid style={gray!25},
	legend style={{nodes={scale=1, transform shape}},draw=black,fill=white,legend cell align=left,inner sep=1pt,row sep = 0pt,at={(0.999,0.62)}},
	xlabel={\large Iteration number},
	ylabel={\large Average sum rate (bps/Hz)},
	ylabel near ticks]
	
	\addplot[line width=1.2pt,solid,color=myblue] table [y=PG2, x=iter,col sep = comma]{convergence2.csv};
	\addlegendentry{\large APG (proposed)}
	
	\addplot[line width=1.2pt,densely dashed,color=myred] table [y=MM2, x=iter,col sep = comma]{convergence2.csv};
	\addlegendentry{\large \cite[Algo.~2]{Zhou20IRS}}
	
	\addplot[line width=1.2pt,solid,color=myblue] table [y=PG1, x=iter,col sep = comma]{convergence2.csv};
	
	\addplot[line width=1.2pt,densely dashed,color=myred] table [y=MM1, x=iter,col sep = comma]{convergence2.csv};
	
	\node[align=center,fill=white,inner sep=3pt] at (axis cs: 750, 10.5) {\large $P_{\mathrm{t}} = 30$~dBm};
	\draw[->,>=latex,line width = 1pt,color=black] (axis cs:450, 10.5) -- (axis cs:250, 11);
	\draw[line width = 1pt, color=black] (axis cs:200, 11.5) arc (130:-250: 0.14cm and  0.95cm);
	
	\node[align=center,fill=white,inner sep=3pt] at (axis cs: 750, 3) {\large $P_{\mathrm{t}} = 10$~dBm};
	\draw[->,>=latex,line width = 1pt,color=black] (axis cs:460, 3) -- (axis cs:300, 2.4);
	\draw[line width = 1pt, color=black] (axis cs:250, 2) arc (180:-155: 0.14cm and  0.4cm);	
	
	\end{axis}
	\end{tikzpicture}
}\vskip-0.1in\caption{Convergence of the APG and MM algorithms for $M=100$ and $G=K_{g}=3,\forall g\in\mathcal{G}$.}
\label{fig:convergence}
\par\end{center}%
\end{minipage}\hfill{}%
\begin{minipage}[t]{0.32\linewidth}%
\begin{center}
\newcommand{\vasymptote}[2][]{
	\draw [densely dashed,#1] ({rel axis cs:0,0} -| {axis cs:#2,0}) -- ({rel axis cs:0,1} -| {axis cs:#2,0});
}
\pgfplotsset{every tick label/.append style={font=\large}}
\resizebox{0.98\columnwidth}{0.75\columnwidth}{
	\begin{tikzpicture}
	\begin{axis}[
	ymin=0,ymax=16,xmin=10,xmax=30,
	xtick={10,15,20,25,30},
	ytick={0,3,6,9,12,15},
	grid=both,
	minor grid style={gray!25},
	major grid style={gray!25},
	legend style={{nodes={scale=1, transform shape}},draw=black,fill=white,legend cell align=left,inner sep=1pt,row sep = 0pt,at={(0.56,1)}},
	xlabel={\large Transmit power ($P_t$)},
	ylabel={\large Average sum rate (bps/Hz)},
	ylabel near ticks,
	]
	
	\addplot[line width=1.2pt,solid,color=myblue,mark=diamond, mark size = 3pt] table [y=PG2, x=Pt,col sep = comma]{RateVsPt.csv};
	\addlegendentry{\large APG (proposed)}
	
	\addplot[line width=1.2pt,densely dashed,color=myred,mark=o,mark size = 2pt] table [y=MM2, x=Pt,col sep = comma]{RateVsPt.csv};
	\addlegendentry{\large MM \cite[Algo.~2]{Zhou20IRS}}
	
	\addplot[line width=1.2pt,solid,color=myblue,mark=diamond, mark size = 3pt] table [y=PG1, x=Pt,col sep = comma]{RateVsPt.csv};
	
	\addplot[line width=1.2pt,densely dashed,color=myred,mark=o, mark size = 2pt] table [y=MM1, x=Pt,col sep = comma]{RateVsPt.csv};
	
	\node[align=center,fill=white,inner sep=3pt] at (axis cs: 25, 4) {\large $N=4$};
	\draw[->,>=latex,line width = 1pt,color=black] (axis cs:23, 4) -- (axis cs:20.7, 5.4);
	\draw[line width = 1pt, color=black] (axis cs:20, 6.5) arc (120:-120: 0.14cm and  0.4cm);
	
	\node[align=center,fill=white,inner sep=3pt] at (axis cs: 14.5, 10.2) {\large $N=12$};
	\draw[->,>=latex,line width = 1pt,color=black] (axis cs:17, 10) -- (axis cs:19.3, 8.6);
	\draw[line width = 1pt, color=black] (axis cs:20, 9.5) arc (60:290: 0.14cm and  0.4cm);

	\node[align=center,fill=white,inner sep=3pt,fill=none] at (axis cs: 28.4, 10) {\color{mygreen}\small $23\%$};
	\draw[<->,>=latex,line width = 1pt,color=black] (axis cs:29.5, 11.7) -- (axis cs:29.5, 9.5);

	\node[align=center,fill=white,inner sep=3pt,fill=none] at (axis cs: 28.4, 13.4) {\color{mygreen}\small $19\%$};
	\draw[<->,>=latex,line width = 1pt,color=black] (axis cs:29.5, 15.2) -- (axis cs:29.5, 12.6);


	\end{axis}
	\end{tikzpicture}
}\vskip-0.1in\caption{Average sum rate versus the transmit power for $G=K_{1}=K_{2}=2$
and $M=100$.}
\label{fig:RatevsPt}
\par\end{center}%
\end{minipage}\hfill{}%
\begin{minipage}[t]{0.32\linewidth}%
\begin{center}
\newcommand{\vasymptote}[2][]{
	\draw [densely dashed,#1] ({rel axis cs:0,0} -| {axis cs:#2,0}) -- ({rel axis cs:0,1} -| {axis cs:#2,0});
}
\pgfplotsset{every tick label/.append style={font=\large}}
\resizebox{0.98\columnwidth}{0.75\columnwidth}{
	\begin{tikzpicture}
	\begin{axis}[
	ymin=8,ymax=19,xmin=25,xmax=400,
	xtick={25,100,225,400},
	xticklabels={$5^2$,$10^2$,$15^2$,$20^2$},
	ytick={8,10,12,14,16,18},
	grid=both,
	minor grid style={gray!25},
	major grid style={gray!25},
	legend style={{nodes={scale=1, transform shape}},draw=black,fill=white,legend cell align=left,inner sep=1pt,row sep = 0pt,at={(0.997,0.185)}},
	xlabel={\large Number of IRS elements ($M$)},
	ylabel={\large Average sum rate (bps/Hz)},
	ylabel near ticks,
	]
	
	\addplot[line width=1.2pt,solid,color=myblue,mark=diamond, mark size = 3pt] table [y=PG2, x=M,col sep = comma]{RateVsM.csv};
	\addlegendentry{\large APG (proposed)}
	
	\addplot[line width=1.2pt,densely dashed,color=myred,mark=o, mark size = 2pt] table [y=MM2, x=M,col sep = comma]{RateVsM.csv};
	\addlegendentry{\large MM \cite[Algo.~2]{Zhou20IRS}}
	
	\addplot[line width=1.2pt,solid,color=myblue,mark=diamond, mark size = 3pt] table [y=PG1, x=M,col sep = comma]{RateVsM.csv};
	
	\addplot[line width=1.2pt,densely dashed,color=myred,mark=o, mark size = 2pt] table [y=MM1, x=M,col sep = comma]{RateVsM.csv};
		
	\node[align=center,fill=white,inner sep=3pt] at (axis cs: 180, 14.7) {\large $N=12$};
	\draw[->,>=latex,line width = 1pt,color=black] (axis cs:145, 14.6) -- (axis cs:100, 15.5);
	\draw[->,>=latex,line width = 1pt,color=black] (axis cs:145, 14.2) -- (axis cs:100, 13);
	
	\node[align=center,fill=white,inner sep=3pt] at (axis cs: 180, 11) {\large $N=4$};
	\draw[->,>=latex,line width = 1pt,color=black] (axis cs:145, 11) -- (axis cs:100, 12.);
	\draw[->,>=latex,line width = 1pt,color=black] (axis cs:145, 10.7) -- (axis cs:100, 10);

%
	\node[align=center,fill=white,inner sep=3pt] at (axis cs: 370, 12) {\color{mygreen}\small $51\%$};
	\draw[<->,>=latex,line width = 1pt,color=black] (axis cs:390, 15.7) -- (axis cs:390, 10.4);
	
	\node[align=center,fill=white,inner sep=3pt] at (axis cs: 360, 16.5) {\color{mygreen}\small $35\%$};
	\draw[<->,>=latex,line width = 1pt,color=black] (axis cs:380, 18.2) -- (axis cs:380, 13.5);
	
	\end{axis}
	\end{tikzpicture}
}\vskip-0.1in\caption{Average sum rate versus $M$ for $G=K_{1}=K_{2}=2$ and $P_{\mathrm{t}}=30$
dBm.}
\label{fig:RatevsM}
\par\end{center}%
\end{minipage}
\end{figure*}

Equipped with \textbf{Theorems~\ref{thm:Grad-f} }and\textbf{~\ref{thm:Grad-theta}},
we are now in a position to described the proposed APG algorithm to
obtain a stationary solution to~\eqref{eq:approxproblem}. The APG
algorithm is summarized in~\textbf{Algorithm~\ref{alg:projGrad}},
where $\mathbf{f}^{(n)}$ and $\boldsymbol{\theta}^{(n)}$ denote
the (stacked) transmit beamformers and IRS phase shifts in the $n$-th
iteration, respectively, and $\alpha_{\mathbf{f}}$ and $\alpha_{\boldsymbol{\theta}}$
denote the step size corresponding to $\mathbf{f}$ and $\boldsymbol{\theta}$,
respectively. Given the previous iteration $\mathbf{f}^{(n-1)}$,
to increase the objective, we move along its gradient direction with
a step size $\alpha_{\mathbf{f}}$ to obtain $\hat{\mathbf{f}}^{(n)}\triangleq\mathbf{f}^{(n-1)}+\alpha_{\mathbf{f}}\nabla_{\mathbf{f}}\tilde{\mathcal{R}}_{\mathrm{sum}}\big(\mathbf{f}^{(n-1)},\boldsymbol{\theta}^{(n-1)}\big)\big)$
and then project $\hat{\mathbf{f}}^{(n)}$onto $\mathcal{F}$ to obtain
$\mathbf{f}^{(n)}$, which is given by
\begin{equation}
\mathbf{f}^{(n)}=\Pi_{\mathcal{F}}\big(\hat{\mathbf{f}}^{(n)}\big)=\sqrt{P_{\mathrm{t}}}\hat{\mathbf{f}}^{(n)}\big/\max\big\{\Vert\hat{\mathbf{f}}^{(n)}\Vert,\sqrt{P_{\mathrm{t}}}\big\}.\label{eq:projF}
\end{equation}
In the same way, for a given $\hat{\boldsymbol{\theta}}^{(n)}=\big[\hat{\theta}_{1}^{(n)},\hat{\theta}_{2}^{(n)},\ldots,\hat{\theta}_{M}^{(n)}\big]\trans$,
its projection onto $\varTheta$, i.e., $\Pi_{\varTheta}\big(\hat{\boldsymbol{\theta}}^{(n)}\big)$,
is given by $\boldsymbol{\theta}^{(n)}=\big[\theta_{1}^{(n)},\theta_{2}^{(n)},\ldots,\theta_{M}^{(n)}\big]\trans$,
where 
\begin{equation}
\theta_{m}^{(n)}=\left\{ \negthickspace\negthickspace\begin{array}{cc}
\hat{\theta}_{m}^{(n)}/|\hat{\theta}_{m}^{(n)}|, & \negthickspace\negthickspace\mathrm{if}\ |\hat{\theta}_{m}^{(n)}|\neq0\\
\exp(j\phi),\phi\in[0,2\pi), & \negthickspace\negthickspace\negthickspace\mathrm{otherwise}
\end{array}\negthickspace\negthickspace,\forall m\in\mathcal{M}.\right.\label{eq:projTheta}
\end{equation}

It is noteworthy that appropriate values of $\alpha_{\mathbf{f}}$
and $\alpha_{\boldsymbol{\theta}}$ can be obtained using a \textit{backtracking
line search} scheme, based on Armijo\textendash Goldstein condition~\cite{LineSearch}.
Also, the proof of convergence of~\textbf{Algorithm~\ref{alg:projGrad}}
follows from~\cite{Convergence}.

\subsection{Complexity Analysis}

We now present a detailed complexity analysis of the proposed APG
algorithm, where we count the required number of complex-valued multiplications
in each iteration in~\textbf{Algorithm~\ref{alg:projGrad}.} In
particular, we show that the proposed algorithm has a complexity that
grows linearly with the number of reflecting elements, which is a
notable reduction compared to the methods in~\cite{Zhou20IRS}. For
the sake of tractability, in this section, we assume an equal number
of users in each group and define $\bar{K}\triangleq K_{i}=K/G,\forall i\in\mathcal{G}$
as the number of users per group. It is easy to note that the computational
complexity of~\textbf{Algorithm~\ref{alg:projGrad}} is dominated
by those associated with projected gradient steps~3 and 4 in~\textbf{Algorithm~\ref{alg:projGrad}}.

First, we analyze the complexity of $\nabla_{\mathbf{f}}\tilde{\mathcal{R}}_{\mathrm{sum}}(\mathbf{f},\boldsymbol{\theta};\tau)$,
which is mostly due to the associated computational complexities for
computing $\nabla_{\mathbf{f}_{i}}R_{k,i}(\mathbf{f},\boldsymbol{\theta})$
and $\nabla_{\mathbf{f}_{i}}R_{k,\ell}(\mathbf{f},\boldsymbol{\theta})$
(see~\eqref{eq:gradFclosed}). The complexity of computing $\mathbf{z}_{k,i}$
is of the order of $\mathcal{O}(MN)$ and computing $K$ such terms
requires $\mathcal{O}(KMN)$ multiplications. The complexities associated
with computing $\mathbf{z}_{k,i}\mathbf{f}_{i}$, $(\mathbf{z}_{k,i})\herm\mathbf{z}_{k,i}\mathbf{f}_{i}$
and $\sum_{g\in\mathcal{G}}|\mathbf{z}_{k,i}\mathbf{f}_{g}|$ are
then given by $\mathcal{O}(N)$, $\mathcal{O}(N)$ and $\mathcal{O}(GN)$.
Therefore, it is clear from~\eqref{eq:A2} that the computational
complexity of $\nabla_{\mathbf{f}_{i}}R_{k,i}(\mathbf{f},\boldsymbol{\theta})$
is $\mathcal{O}(GN)$. Analogously, the computational complexity associated
with $\nabla_{\mathbf{f}_{i}}R_{k,\ell}(\mathbf{f},\boldsymbol{\theta})$
is given by $\mathcal{O}(GN)$. From~\eqref{eq:gradFclosed}, it
is then straightforward to see that the complexity of computing $\nabla_{\mathbf{f}_{i}}\tilde{\mathcal{R}}_{\mathrm{sum}}(\mathbf{f},\boldsymbol{\theta})$
is $\mathcal{O}(\bar{K}GN+(G-1)\bar{K}GN)=\mathcal{O}(KGN)$, and
that for computing $\nabla_{\mathbf{f}}\tilde{\mathcal{R}}_{\mathrm{sum}}(\mathbf{f},\boldsymbol{\theta};\tau)$
is $\mathcal{O}(G^{2}KN)$. Note that the complexity of obtaining
appropriate value of $\alpha_{\mathbf{f}}$ and that of $\Pi_{\mathcal{F}}(\cdot)$
will be negligible, and therefore, the per-iteration complexity of
step~3 in~\textbf{Algorithm~\ref{alg:projGrad}} is equal to $\mathcal{O}(KNM+G^{2}KN)$.

Next, in order to estimate the complexity associated with the computation
of $\nabla_{\boldsymbol{\theta}}\tilde{\mathcal{R}}_{\mathrm{sum}}(\mathbf{f},\boldsymbol{\theta})$,
one needs to count the number of complex-valued multiplications required
to compute $\nabla_{\boldsymbol{\theta}}|\mathbf{z}_{k,g}\mathbf{f}_{\jmath}|^{2}$
and $\nabla_{\boldsymbol{\theta}}|\mathbf{z}_{k,g}\mathbf{f}_{\ell}|^{2}$
(see~\eqref{eq:gradThetaClosed}). As given in~\eqref{eq:B3}, $\nabla_{\boldsymbol{\theta}}|\mathbf{z}_{k,g}\mathbf{f}_{\jmath}|^{2}=\vecd\big\{\big(\hat{\mathbf{h}}_{k,g}\big)\herm\mathbf{z}_{k,g}\mathbf{f}_{\jmath}\mathbf{f}_{\jmath}\herm\mathbf{H}_{\mathrm{ts}}\herm\big\}$,
and since we have already computed $\mathbf{z}_{k,g}\mathbf{f}_{\jmath}$
(see discussions in the preceding paragraph), the complexity of computing
$\mathbf{z}_{k,g}\mathbf{f}_{\jmath}\mathbf{f}_{\jmath}\herm\mathbf{H}_{\mathrm{ts}}\herm$
is given by $\mathcal{O}(MN+M)$. Now since we only need to compute
the diagonal elements of $\big(\hat{\mathbf{h}}_{k,g}\big)\herm\mathbf{z}_{k,g}\mathbf{f}_{\jmath}\mathbf{f}_{\jmath}\herm\mathbf{H}_{\mathrm{ts}}\herm$,
the complexity associated with $\nabla_{\boldsymbol{\theta}}|\mathbf{z}_{k,g}\mathbf{f}_{\jmath}|^{2}$
is given by $\mathcal{O}(MN+2M)$. Similarly, the computational complexity
of $\nabla_{\boldsymbol{\theta}}|\mathbf{z}_{k,g}\mathbf{f}_{\ell}|^{2}$
is given by $\mathcal{O}(MN+2M)$. Therefore, the complexity associated
with $\nabla_{\boldsymbol{\theta}}\tilde{\mathcal{R}}(\mathbf{f},\boldsymbol{\theta})$
is $\mathcal{O}(G^{2}\bar{K}(MN+2M)+G(G-1)\bar{K}(MN+2M))\approx\mathcal{O}(MKGN)$.
The complexity of backtracking line search to obtain appropriate value
of $\alpha_{\boldsymbol{\theta}}$ and that for $\Pi_{\varTheta}(\cdot)$
is comparatively very small, and can therefore be neglected. Hence,
the total complexity associated with step~4 in~\textbf{Algorithm~\ref{alg:projGrad}}
will be the same as that of the $\nabla_{\boldsymbol{\theta}}\tilde{\mathcal{R}}(\mathbf{f},\boldsymbol{\theta})$.

From the discussions presented above, we can write the overall per-iteration
complexity of~\textbf{Algorithm~\ref{alg:projGrad}} as
\begin{equation}
\mathcal{O}(MKGN+G^{2}KN).\label{eq:FinalCompelxity}
\end{equation}
Since in a practical IRS-MGMC system, the number of IRS tiles is likely
much larger than the number of transmit antennas, total number of
users, or the total number of groups, i.e., $M\gg\max\{N,K,G\}$,
the overall per-iteration complexity of~\textbf{Algorithm~\ref{alg:projGrad}}
can be approximated by $\mathcal{O}(MKGN)$, which is linear w.r.t.
the number of IRS tiles. Recall that as discussed in Sec.~\ref{sec:System-Model},
the complexity of the MM algorithm~\cite{Zhou20IRS} has a cubic
growth w.r.t. the number of tiles at the IRS.

\section{\label{sec:Numerical-Results}Numerical Results}

\begin{figure}[t]
\begin{centering}
\newcommand{\vasymptote}[2][]{
	\draw [densely dashed,#1] ({rel axis cs:0,0} -| {axis cs:#2,0}) -- ({rel axis cs:0,1} -| {axis cs:#2,0});
}
\pgfplotsset{every tick label/.append style={font=\large}}
\resizebox{0.8\columnwidth}{!}{
	\begin{tikzpicture}
	\begin{axis}[
	ymode=log,
	log basis y={10},
	ymin=0.1,ymax=1100,xmin=25,xmax=400,
	xtick={25,100,225,400},
	xticklabels={$5^2$,$10^2$,$15^2$,$20^2$},
	grid=both,
	minor grid style={draw=none},
	major grid style={gray!25},
	legend style={{nodes={scale=1, transform shape}},draw=black,fill=white,legend cell align=left,inner sep=1pt,row sep = 0pt,at={(1,0.6)}},
	xlabel={Number of IRS elements ($M$)},
	ylabel={Average run time (s)},
	ylabel near ticks,
	]

	\addplot[line width=1.2pt,dashed,color=myred,mark=o, mark size = 2pt] table [y=MM1, x=M,col sep = comma]{runtime.csv};
	\addlegendentry{MM \cite[Algo.~2]{Zhou20IRS}: $N = 4$}	
	
	\addplot[line width=1.2pt,dotted,color=myred,mark=square*, mark size = 2pt] table [y=MM2, x=M,col sep = comma]{runtime.csv};
	\addlegendentry{MM \cite[Algo.~2]{Zhou20IRS}: $N = 12$}
	
	\addplot[line width=1.2pt,dashdotted,color=myblue,mark=triangle*, mark size = 3pt] table [y=PG2, x=M,col sep = comma]{runtime.csv};
	\addlegendentry{APG (proposed): $N = 12$}
	
	\addplot[line width=1.2pt,solid,color=myblue,mark=diamond, mark size = 3pt] table [y=PG1, x=M,col sep = comma]{runtime.csv};
	\addlegendentry{APG (proposed): $N = 4$}

%

	\end{axis}
	\end{tikzpicture}
}
\par\end{centering}
\caption{Average run-time comparison between the proposed APG and the MM algorithms
for $G=K_{1}=K_{2}=2$ and $P_{\mathrm{t}}=30$~dBm.}
\label{fig:runTime}
\end{figure}
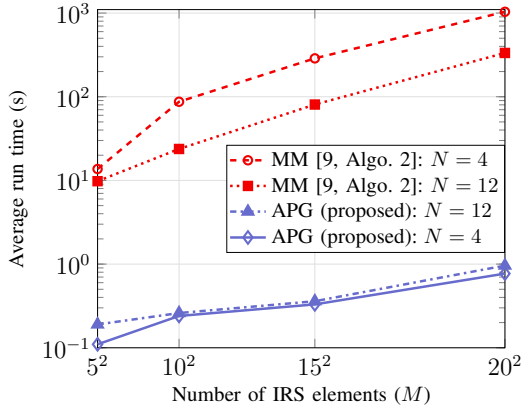

In this section, we present the results of numerical experiments to
evaluate the performance of the system under consideration. It is
assumed that the uniform linear array at the transmitter is centered
at ($0$ m, $20$ m, $10$ m), whereas the uniform planar array at
the IRS is centered at ($30$ m, $0$ m, $5$ m). On the other hand,
the users are assumed to be uniformly distributed in a circular area
of radius $20\ \mathrm{m}$, centered at ($350$ m, $50$ m, $2$
m). The center frequency of the carrier wave is set to $2$ GHz. The
distance between the adjacent antennas at the transmitter, and that
between the adjacent tiles at the IRS is considered to be $\lambda/2$,
with $\lambda\ (=0.15\ \mathrm{m})$ being the carrier wavelength.
On the other hand, the minimum distance between the users is assumed
to be equal to $2\lambda$. The distance-dependent path loss and the
Rician-distributed small-scale fading between the nodes are modeled
following the arguments in~\cite[Sec.~VI]{StefanMIMOBC}. The noise
power spectral density is equal to $-174$ dBm/Hz, and the total available
bandwidth is $10$ MHz. In Figs.~\ref{fig:convergence}\textendash \ref{fig:runTime},
the average achievable sum rate/average run time is shown for $100$
channel realizations. Also, we consider $\tau=50$, and the tolerance
for convergence is set to $10^{-5}$.

In Fig.~\ref{fig:convergence}, we compare the convergence of the
proposed APG algorithm with that of the baseline MM algorithm. It
can be observed from the figure that the proposed APG algorithm outperforms
the baseline MM algorithm, and the difference between the performance
of the two algorithms increases with increasing transmit power. This
is due to the reason that the tightness of the bounds used in~\cite[Algo.~2]{Zhou20IRS}
for the MM algorithm depends on the transmit power and the involved
channels. In particular, if the transmit power is small, then the
bounds are relatively tight but are not so for a high transmit power
(cf. \cite[Appendix D]{Zhou20IRS}). The consequence is that the progress
made in each iteration of the MM method is very small. Thus, even
after a very large number of iterations, the MM algorithm still does
not reaches a full convergence. It can also be observed that the proposed
APG method requires quite many iterations to converge for high transmit
power since in this case, the Lipschitz constant of the gradient of
the objective is large, which forces the step size in each iteration
to be small. However, since the per-iteration complexity is very small,
the overall run-time for the APG algorithm is much lesser than that
of the MM algorithm. We discuss the average run time for both the
algorithms later in~Fig.~\ref{fig:runTime}.

Next, in Fig.~\ref{fig:RatevsPt}, we demonstrate the effect of transmit
power on the achievable sum rate, and also the advantage of deploying
more transmit antennas in the considered system. It can be observed
from the figure that the proposed APG algorithm outperforms the MM
algorithm for all considered scenarios. The performance difference
between the two algorithms increases with increasing values of $P_{\mathrm{t}}$
due to the same reason as explained for~Fig.~\ref{fig:convergence}.
We also remark that the sum rate increases with an increase in the
number of transmit antennas due to the multiplexing gains of MISO
systems. The gains compared to the MM algorithms are clearly indicated
in Fig~\ref{fig:RatevsPt}.

In Fig.~\ref{fig:RatevsM}, we compare the sum rate performance of
the proposed APG algorithm with that of the MM algorithm for an increasing
number of IRS tiles. From the figure, it can be noticed that increasing
the number of IRS tiles improves the achievable sum rate as a larger
number of IRS tiles enables the IRS to perform highly-focused beamforming
to maximize the achievable sum rate. Next, we observe that when the
number of IRS tiles is increased, the rate of increase in the achievable
sum rate is much higher in the case of the proposed APG method than
the baseline MM method, where the gains compared to the MM algorithms
are clearly marked in Fig~\ref{fig:RatevsM}.

In Fig.~\ref{fig:runTime}, we show the average rum time comparison
between the proposed APG algorithm and the MM algorithm\footnote{We are thankful to the authors of \cite{Zhou20IRS} for sending us
the code for their proposed MM methods for comparison.}. As described earlier, the complexity of the proposed APG algorithm
grows linearly, in comparison to the cubic rate of increase in the
complexity of the MM algorithm w.r.t. $M$. This result is in-line
with that shown in Fig.~\ref{fig:runTime}, where for large value
of $M$ (say 400), the APG algorithm is $1000$ times faster than
the MM algorithm. The average run time of the APG algorithm increases
when the number of transmit antennas increases, because the complexity
of the proposed algorithm also grows with $N$ (see~\eqref{eq:FinalCompelxity}).
Interestingly enough, for fixed values of $M,K$ and $G$, the average
run time of the MM algorithm decreases with increasing $N$. This
occurs because, for a large value of $N$, the system has more degrees
of freedom, which results in faster convergence of the algorithm (compared
to the case when $N$ is small). This in turn reduces the average
run time for the MM algorithm when $N$ is large.

\section{Conclusion}

In this paper, we have considered the problem of sum rate maximization
for an IRS-assisted multigroup multicast MISO system. In order to
jointly design the optimal transmit beamformer and IRS phase shifts,
we proposed a low-complexity alternating projected gradient method,
that outperformed the benchmark schemes both in terms of performance
and complexity. The complexity analysis confirmed that the complexity
of the proposed algorithm increases linearly with the number of reflecting
elements at the IRS, which is the best-known complexity result so
far for such IRS-assisted systems. Extensive numerical results were
provided to insight into the achievable rate performance of the IRS-MGMC
system for different system parameters.

\appendices{}

\section{\label{sec:Proof-1}Proof of Theorem~\ref{thm:Grad-f}}

Using~\eqref{eq:approxsystemrate}, it is easy to note that \small{
\begin{align}
 & \nabla_{\mathbf{f}_{i}}\tilde{\mathcal{R}}_{\mathrm{sum}}(\mathbf{f},\boldsymbol{\theta})\nonumber \\
= & \sum_{g\in\mathcal{G}}\frac{\sum\nolimits _{k\in\mathcal{K}_{g}}\bigg[\exp\big\{-\tau R_{k,g}(\mathbf{f},\boldsymbol{\theta})\big\}\nabla_{\mathbf{f}_{i}}R_{k,g}(\mathbf{f},\boldsymbol{\theta})\bigg]}{\sum\nolimits _{k\in\mathcal{K}_{g}}\exp\big\{-\tau R_{k,g}(\mathbf{f},\boldsymbol{\theta})\big\}}\nonumber \\
= & \frac{\sum\nolimits _{k\in\mathcal{K}_{i}}\bigg[\exp\big\{-\tau R_{k,i}(\mathbf{f},\boldsymbol{\theta})\big\}\nabla_{\mathbf{f}_{i}}R_{k,i}(\mathbf{f},\boldsymbol{\theta})\bigg]}{\sum\nolimits _{k\in\mathcal{K}_{i}}\exp\big\{-\tau R_{k,i}(\mathbf{f},\boldsymbol{\theta})\big\}}\nonumber \\
+ & \!\!\sum_{\ell\in\mathcal{G}\setminus\{i\}}\!\!\frac{\sum\nolimits _{k\in\mathcal{K}_{\ell}}\bigg[\!\!\exp\big\{\!-\!\tau R_{k,\ell}(\mathbf{f},\boldsymbol{\theta})\big\}\nabla_{\mathbf{f}_{i}}R_{k,\ell}(\mathbf{f},\boldsymbol{\theta})\!\bigg]}{\sum\nolimits _{k\in\mathcal{K}_{\ell}}\exp\big\{-\tau R_{k,\ell}(\mathbf{f},\boldsymbol{\theta})\big\}}.\label{eq:A1}
\end{align}
}Thus, to derive $\nabla_{\mathbf{f}_{i}}\tilde{\mathcal{R}}_{\mathrm{sum}}(\mathbf{f},\boldsymbol{\theta})$
we need to find $\nabla_{\mathbf{f}_{i}}R_{k,i}(\mathbf{f},\boldsymbol{\theta})$
and $\nabla_{\mathbf{f}_{i}}R_{k,\ell}(\mathbf{f},\boldsymbol{\theta})$
for $l\in\mathcal{G}\setminus\{i\}$. To this end using~\eqref{eq:userrate},
we have \small{
\begin{align}
 & \nabla_{\mathbf{f}_{i}}R_{k,i}(\mathbf{f},\boldsymbol{\theta})=\nabla_{\mathbf{f}_{i}}\ln\bigg(1+\sum_{g\in\mathcal{G}}|\mathbf{z}_{k,i}\mathbf{f}_{g}|^{2}\bigg)\nonumber \\
 & \qquad\qquad\qquad\qquad-\underbrace{\nabla_{\mathbf{f}_{i}}\ln\bigg(1+\sum_{\ell\in\mathcal{G}\setminus\{i\}}|\mathbf{z}_{k,i}\mathbf{f}_{\ell}|^{2}\bigg)}_{=\boldsymbol{0}}\nonumber \\
= & \frac{\nabla_{\mathbf{f}_{i}}\big(\mathbf{z}_{k,i}\mathbf{f}_{i}\mathbf{f}_{i}\herm\mathbf{z}_{k,i}\herm\big)}{1+\sum_{g\in\mathcal{G}}|\mathbf{z}_{k,i}\mathbf{f}_{g}|^{2}}\overset{(\mathrm{a})}{=}\frac{\mathbf{z}_{k,i}\herm\mathbf{z}_{k,i}\mathbf{f}_{i}}{1+\sum_{g\in\mathcal{G}}|\mathbf{z}_{k,i}\mathbf{f}_{g}|^{2}},\label{eq:A2}
\end{align}
}where $(\mathrm{a})$ follows from~\cite[Table~4.3]{DerivativeBook}.
Following similar steps, it can be shown that\small{
\begin{align}
 & \nabla_{\mathbf{f}_{i}}R_{k,\ell}(\mathbf{f},\boldsymbol{\theta})\!=\!\bigg(\frac{\mathbf{z}_{k,\ell}\herm\mathbf{z}_{k,\ell}\mathbf{f}_{i}}{1+\sum_{g\in\mathcal{G}}|\mathbf{z}_{k,\ell}\mathbf{f}_{g}|^{2}}-\frac{\mathbf{z}_{k,\ell}\herm\mathbf{z}_{k,\ell}\mathbf{f}_{i}}{1+\sum_{\jmath\in\mathcal{G}\setminus\{\ell\}}|\mathbf{z}_{k,\ell}\mathbf{f}_{\jmath}|^{2}}\bigg).\label{eq:A3}
\end{align}
}Inserting $\nabla_{\mathbf{f}_{i}}R_{k,i}(\mathbf{f},\boldsymbol{\theta})$
and $\nabla_{\mathbf{f}_{i}}R_{k,\ell}(\mathbf{f},\boldsymbol{\theta})$
from~\eqref{eq:A2} and~\eqref{eq:A3}, respectively, into~\eqref{eq:A1}
gives $\nabla_{\mathbf{f}_{i}}\tilde{\mathcal{R}}_{\mathrm{sum}}(\mathbf{f},\boldsymbol{\theta})$
expressed in~\eqref{eq:gradFclosed}, which completes the proof.

\section{\label{sec:Proof-2}Proof of Theorem \ref{thm:Grad-theta}}

From~\eqref{eq:approxsystemrate}, it follows that \small{
\begin{align}
 & \nabla_{\boldsymbol{\theta}}\tilde{R}_{\mathrm{sum}}(\mathbf{f},\boldsymbol{\theta})\nonumber \\
= & \sum_{g\in\mathcal{G}}\frac{\sum_{k\in\mathcal{K}_{g}}\bigg[\exp\big\{-\tau R_{k,g}(\mathbf{f},\boldsymbol{\theta})\big\}\nabla_{\boldsymbol{\theta}}R_{k,g}(\mathbf{f},\boldsymbol{\theta})\bigg]}{\sum_{k\in\mathcal{K}_{g}}\exp\{-\tau R_{k,g}(\mathbf{f},\boldsymbol{\theta})\}}.\label{eq:B1}
\end{align}
}It is clear that we now need to find $\nabla_{\boldsymbol{\theta}}R_{k,g}(\mathbf{f},\boldsymbol{\theta})$,
which is given by \small{
\begin{align}
 & \nabla_{\boldsymbol{\theta}}R_{k,g}(\mathbf{f},\boldsymbol{\theta})=\nabla_{\boldsymbol{\theta}}\ln\bigg(1+\sum_{\jmath\in\mathcal{G}}|\mathbf{z}_{k,g}\mathbf{f}_{\jmath}|^{2}\bigg)\nonumber \\
 & \qquad\qquad\qquad\qquad-\nabla_{\boldsymbol{\theta}}\ln\bigg(1+\sum_{\ell\in\mathcal{G}\setminus\{\jmath\}}|\mathbf{z}_{k,g}\mathbf{f}_{\ell}|^{2}\bigg)\nonumber \\
= & \frac{\sum_{\jmath\in\mathcal{G}}\nabla_{\boldsymbol{\theta}}|\mathbf{z}_{k,g}\mathbf{f}_{\jmath}|^{2}}{1+\sum_{\jmath\in\mathcal{G}}|\mathbf{z}_{k,g}\mathbf{f}_{\jmath}|^{2}}-\frac{\sum_{\ell\in\mathcal{G}\setminus\{\jmath\}}\nabla_{\boldsymbol{\theta}}|\mathbf{z}_{k,g}\mathbf{f}_{\ell}|^{2}}{1+\sum_{\ell\in\mathcal{G}\setminus\{\jmath\}}|\mathbf{z}_{k,g}\mathbf{f}_{\ell}|^{2}}.\label{eq:B2}
\end{align}
}Next, a closed-form expression for $\nabla_{\boldsymbol{\theta}}|\mathbf{z}_{k,g}\mathbf{f}_{\jmath}|^{2}$
can be obtained as follows:\small{
\begin{align}
 & \nabla_{\boldsymbol{\theta}}|\mathbf{z}_{k,g}\mathbf{f}_{\jmath}|^{2}=\nabla_{\boldsymbol{\theta}}\Big(\mathbf{z}_{k,g}\mathbf{f}_{\jmath}\mathbf{f}_{\jmath}\herm\big(\mathbf{z}_{k,g}\big)\herm\Big)\nonumber \\
= & \nabla_{\boldsymbol{\theta}}\Big[\big(\mathbf{h}_{k,g}+\hat{\mathbf{h}}_{k,g}\boldsymbol{\Theta}\mathbf{H}_{\mathrm{ts}}\big)\mathbf{f}_{\jmath}\mathbf{f}_{\jmath}\herm\big\{(\mathbf{h}_{k,g})\herm\!+\!\mathbf{H}_{\mathrm{ts}}\herm\boldsymbol{\Theta}\herm\big(\hat{\mathbf{h}}_{k,g}\big)\herm\big\}\Big]\nonumber \\
\overset{(\mathrm{b})}{=} & \vecd\big\{\big(\hat{\mathbf{h}}_{k,g}\big)\herm\mathbf{z}_{k,g}\mathbf{f}_{\jmath}\mathbf{f}_{\jmath}\herm\mathbf{H}_{\mathrm{ts}}\herm\big\},\label{eq:B3}
\end{align}
}where $(\mathrm{b})$ follows from~\cite[Table~4.3 and eqn.~(6.153)]{DerivativeBook}.
Using~\eqref{eq:B1}-\eqref{eq:B3}, a closed-form expression for
$\nabla_{\boldsymbol{\theta}}\tilde{\mathcal{R}}_{\mathrm{sum}}(\mathbf{f},\boldsymbol{\theta})$
is given by~\eqref{eq:gradThetaClosed}. This concludes the proof.

\bibliographystyle{IEEEtran}
\bibliography{IEEEabrv,bibTex}

\end{document}